\documentclass{JINST}
\let\ifpdf\relax

\usepackage{lineno}
\usepackage{subfigure}
\usepackage{comment}

\newcommand\lum{\ensuremath{\mathrm{cm}^{-2} \mathrm{s}}} 
\newcommand{\neqcm}{\ensuremath{\mathrm{n}_{\mathrm{eq}}/\mathrm{cm}^2}}
\newcommand{\mum}{$\mathrm{\mu}$m}

\newcommand{\degC}{\ensuremath{^{\circ}\mathrm{C}}}

\newcommand{\Sr}{\ensuremath{^{90}}Sr}
\newcommand{\Cd}{\ensuremath{^{109}}Cd}
\newcommand{\Am}{\ensuremath{^{241}}Am}

\graphicspath{{./images/}}

\title{Heavily Irradiated N-in-p Thin Planar Pixel Sensors with and without Active Edges}

\author{S.~Terzo$^a$\thanks{Corresponding author.}~, L.~Andricek$^{b}$, A.~Macchiolo$^a$, H.G.~Moser$^{a,b}$, R.~Nisius$^a$, R.H.~Richter$^{b}$
and P.~Weigell$^a$\\
\llap{$^a$}Max-Planck-Institut f\"ur Physik (Werner-Heisenberg-Institut),\\
  F\"ohringer Ring 6, D-80805 M\"unchen, Germany\\
\llap{$^b$}Max-Planck-Gesellschaft Halbleiterlabor,\\
  Otto Hahn Ring 6, D-81739 M\"unchen, Germany\\
E-mail: \email{Stefano.Terzo@mpp.mpg.de}}

\abstract{We present the results of the characterization of silicon pixel modules employing n-in-p 
planar sensors with an active thickness of 150 \mum{}, produced at MPP/HLL, and 100-200 \mum{} thin active edge sensor devices, produced at VTT in Finland. 
These thin sensors are designed as candidates for the ATLAS pixel detector upgrade to be operated at the HL-LHC, as they ensure radiation hardness at high fluences. They are interconnected to the 
ATLAS FE-I3 and FE-I4 read-out chips.
Moreover, the n-in-p technology only requires a single side processing and thereby it is a cost-effective 
alternative to the n-in-n pixel technology presently employed in the LHC experiments. 
High precision beam test measurements of the hit efficiency have been performed on these 
devices both at the CERN SpS and at DESY, Hamburg. We studied the behavior of these 
sensors at different bias voltages and different beam incident angles up to the maximum one 
expected for the new Insertable B-Layer of ATLAS and for HL-LHC detectors. 
Results obtained with 150 \mum{} thin sensors, assembled with the new ATLAS FE-I4 chip and 
irradiated up to a fluence of 4$\times$10$^{15}$ \neqcm{}{}, show that they are excellent candidates for larger 
radii of the silicon pixel tracker in the upgrade of the ATLAS detector at HL-LHC. In addition, the active edge technology 
of the VTT devices maximizes the active area of the sensor 
and reduces the material budget to suit the requirements for the innermost layers. 
The edge pixel performance of VTT modules has been investigated at beam test experiments and 
the analysis after irradiation up to a fluence of 5$\times$10$^{15}$ \neqcm{}{} has been performed using radioactive sources in the laboratory.}

\begin{document}

\section{Introduction}\label{sec:intro}
The upgrade of the LHC planned for 2022 (HL-LHC) aims at increasing the instantaneous luminosity of the accelerator up to 5$\times$10$^{34}$ \lum{}~\cite{hl-lhc}. To face the higher particle fluence expected, and profit from the huge amount of data that the HL-LHC will deliver, an upgrade of the ATLAS detector will be necessary, which will require a full replacement of the inner pixel detector with new radiation hard devices~\cite{phase2}. The foreseen layout will employ 4 or 5 barrel layers to cover a pseudo-rapidity range of $|\eta|\le$3. The distance of the innermost layer from the proton beam line will be only 3.9 cm. Because of the small radius, overlapping of pixel modules of the innermost layer along the beam direction is not allowed by design, therefore the feasibility of new solutions to maximize the active area of the pixel modules is also currently under study.

\section{Pixel modules for high luminosity collider experiments}\label{sec:concept}
New planar pixel module prototypes have been produced and characterized to fulfill the requirements for the HL-LHC upgrade of the ATLAS inner detector. They are based on the n-in-p pixel technology and employ thin sensors as a solution for radiation hardness. N-in-p pixels have already been proven to be a potentially cost effective alternative to n-in-n devices presently used in ATLAS~\cite{n-in-p,n-in-p-kek}. Moreover, with the active edge concept, where the back implantation is extended to the sides of the sensor, it is possible to maximize the active region of the sensor reducing the space from the last pixel implant to the border and collecting charge even outside the pixel area. Various sensor thicknesses and guard ring designs for active edge devices are investigated and compared in this paper. The final module concept is completed with two main innovations on the read-out chip side: the introduction of a novel SLID interconnection technique to replace the standard bump bonding and the development of through silicon vias to bring the signal to the back side of the chip passing directly through the chip itself~\cite{slid-tsv}. The combination of all these technologies aims at the development of a full four-side buttable module for the innermost layers.

\subsection{The thin pixel production at MPP/HLL}\label{sec:thin}
A production of 150 \mum{} thin n-in-p silicon sensors designed at the Max-Planck-Institut f\"ur Physik has been carried out by the Max-Planck-Gesellschaft Halbleiterlabor (MPG-HLL) on 6 inch wafer of p-type Float Zone (FZ) silicon. The thinning process requires a handle wafer attached to the sensor as mechanical support to ensure the rigidity of the structure~\cite{hll-thin}. The chosen inter-pixel isolation method is p-spray. The sensors have been covered with a BCB layer to avoid discharges between the sensor and the read-out chip at high voltages, and then interconnected at IZM with bump bonding to  ATLAS FE-I4 read-out chips~\cite{fei4}.
Irradiations up to a fluence of 10$^{16}$ \neqcm{} have been performed at the Los Alamos Neutron Science Center (LANSCE), at the TRIGA Mark II research reactor in Ljubljana and at the Compact Cyclotron of the Karlsruher Institut f\"ur Technologie (KIT).

\paragraph{Experimental setup.} Measurements of the sensor efficiency before and after irradiation have been performed in beam test experiments inside the PPS collaboration both at DESY, Hamburg with 4 GeV electrons and at the CERN SpS using 120 GeV pions.
For particle tracking the EUDET telescope is used, which allows to obtain a pointing resolution on the Device Under Test (DUT) as low as 2 \mum{} in the case of high energetic pions~\cite{EUDET,eff-err}.
For testing of irradiated devices the DUTs are cooled with dry-ice to a measured temperature on the sensor between -50 \degC{} and -40 \degC{}. 

\begin{figure*}[tbp] 
\centering
\subfigure[]{\includegraphics[width=0.49\textwidth]{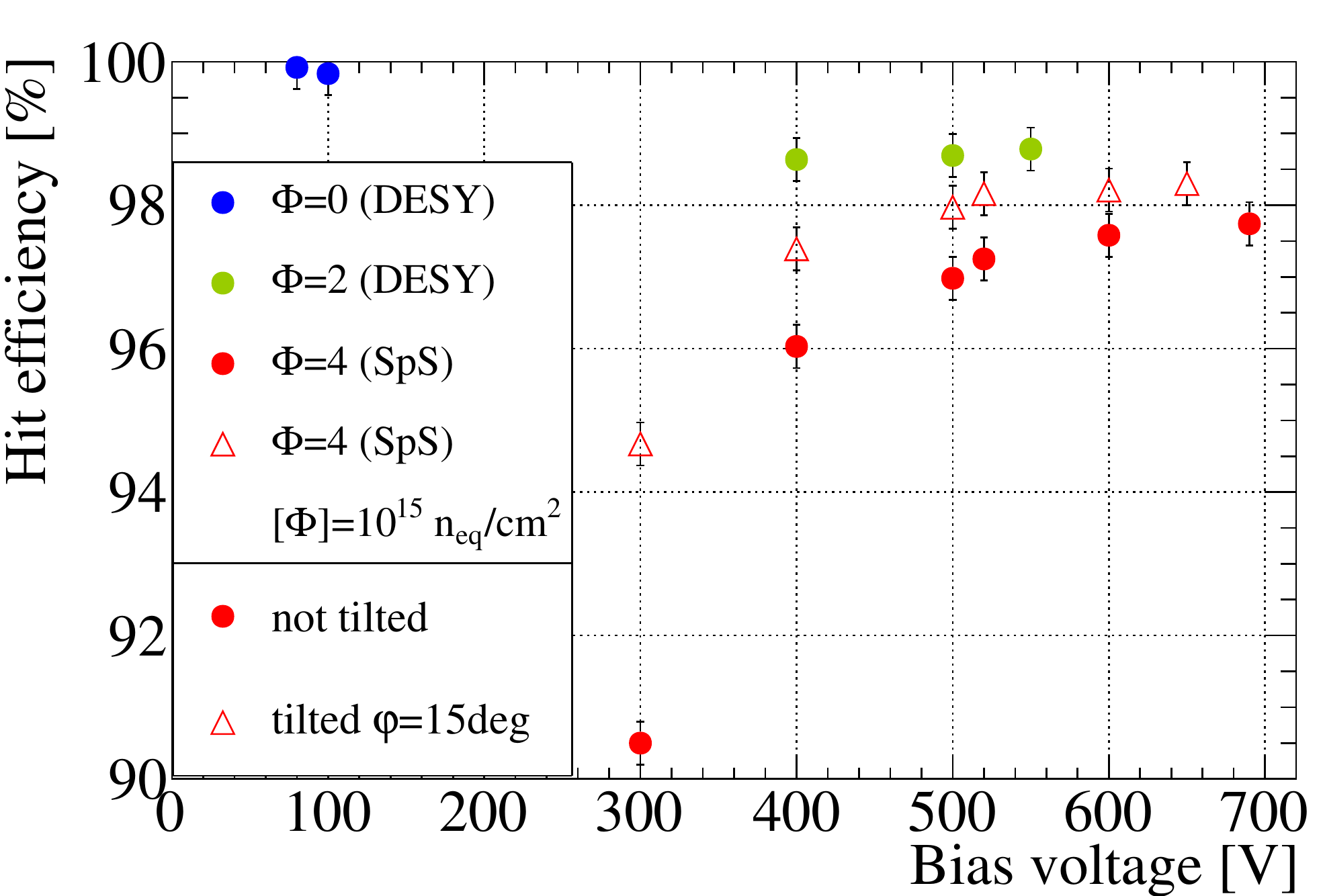}\label{fig:HLL-eff-vs-v_a}}
\subfigure[]{
	\includegraphics[width=0.49\textwidth]{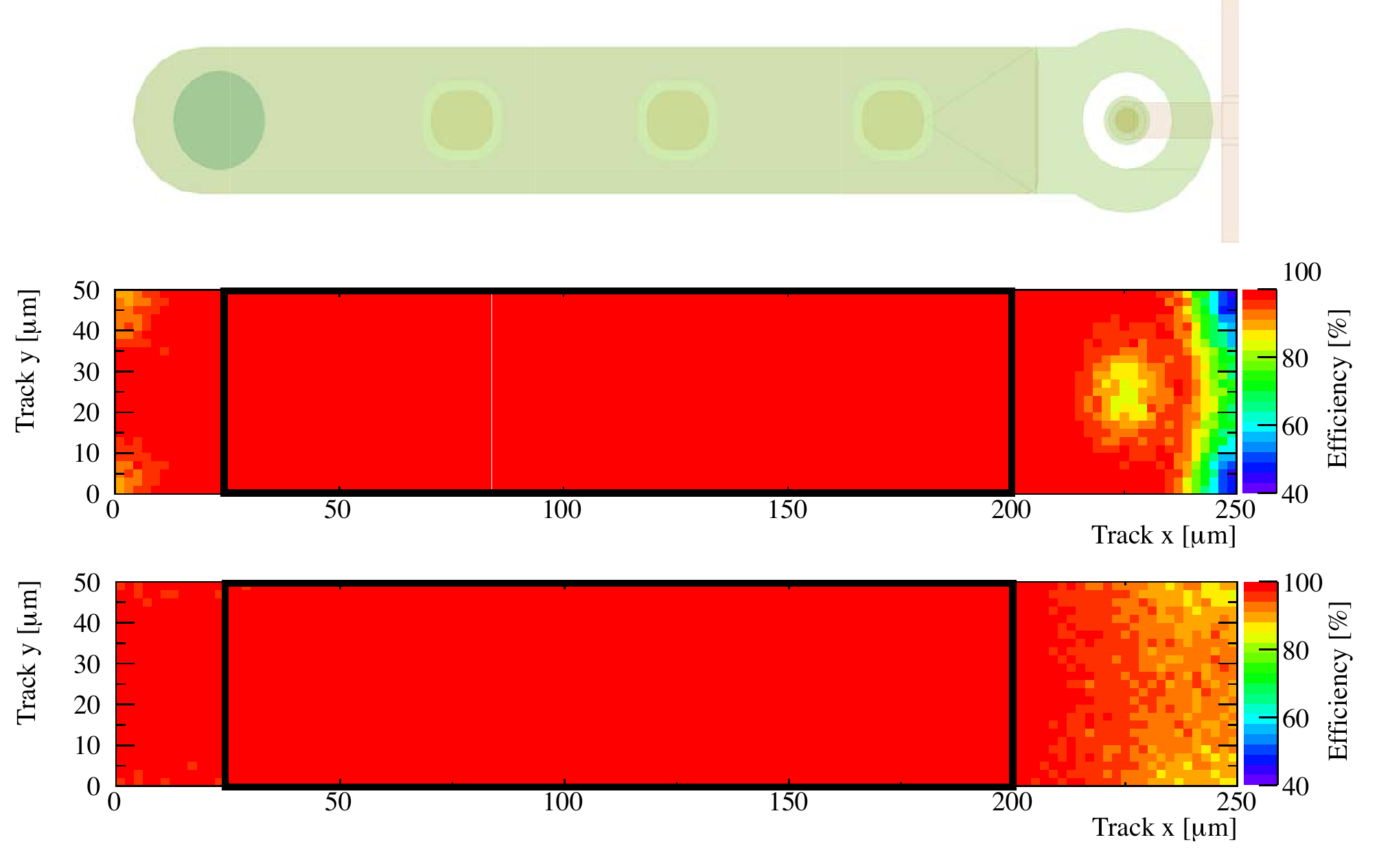}\label{fig:HLL-eff-vs-v_b}
}
\caption{In \protect\subref{fig:HLL-eff-vs-v_a} the hit efficiency as a function of bias voltage is shown for FE-I4 modules with 150 \mum{} thin sensors irradiated to different fluences. At the highest irradiation fluence of 4$\times$10$^{15}$ \neqcm{} the efficency at perpendicular beam incidence is compared to measurements at 15$^{\circ}$ incident angle. Shown in \protect\subref{fig:HLL-eff-vs-v_b} are the respective hit efficiency maps across the pixel surface at the highest measured voltages of 690 V (top) and 650 V (bottom). The maps are obtained associating the hit efficiency to the track crossing position inside a single cell.}

\label{fig:HLL-eff-vs-v}
\end{figure*}

\paragraph{Hit efficiency.} An overview of results of the global hit efficiency at different bias voltages and irradiation fluences is reported in figure~\ref{fig:HLL-eff-vs-v}.
The hit efficiency of the sensor is defined as the ratio of the number of clusters in the DUT associated to a track reconstructed by the telescope to the total number of reconstructed tracks that pass through the active area of the DUT. The impact point of a track is calculated for the middle of the thickness of the active bulk, and a cluster is associated to a track if the track crossed at least one of its pixels. An absolute systematic uncertainty of 0.3\% is associated to all hit efficiency measurements in this paper according to~\cite{eff-err}. For tracks perpendicular to the sensor surface, the hit efficiency drops from 99.9\% before irradiation to 98.8\% after a fluence of 2$\times$10$^{15}$ \neqcm{}. After a fluence of 4$\times$10$^{15}$ \neqcm{} an efficiency of 97.7\% can be still obtained when increasing the bias voltage to $V_{\mathrm{bias}}=690$ V. Figure~\ref{fig:HLL-eff-vs-v_b} shows that the main inefficiency regions inside the single pixel cell after this fluence are located at the punch through area and near the bias rail since these structures are kept at ground potential and not connected to the read-out. This effect is partially overcome for inclined tracks: tilting the device by 15$^{\circ}$ around the axis perpendicular to the short pixel side, a hit efficiency of 98.3\% at $V_{\mathrm{bias}}=650$ V is measured.

\begin{figure*}[tbp] 
\centering
\subfigure[]{
\includegraphics[width=0.4\textwidth]{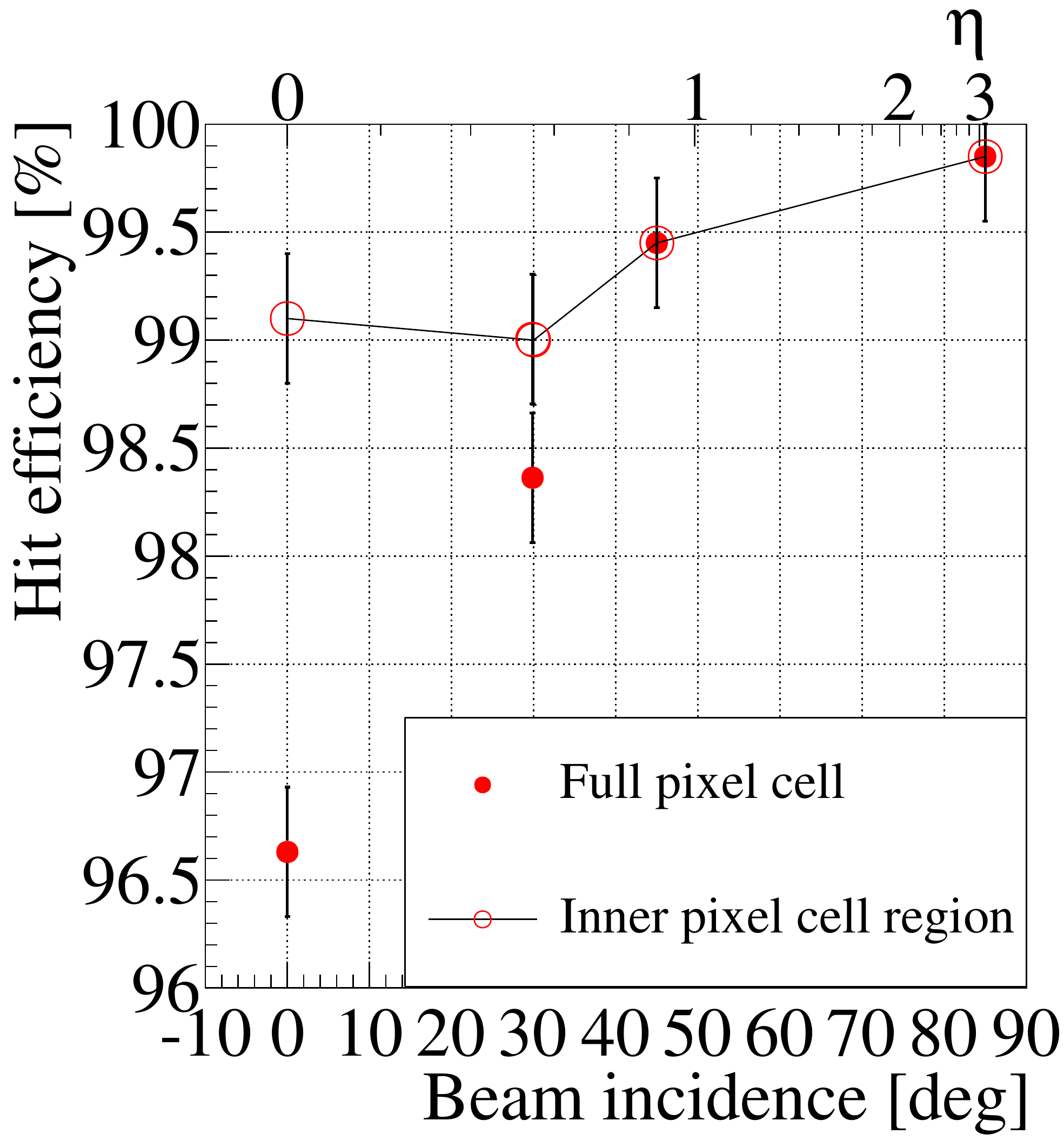}\label{fig:HLL-hieta-a}
}
\subfigure[]{
	\includegraphics[width=0.5\textwidth]{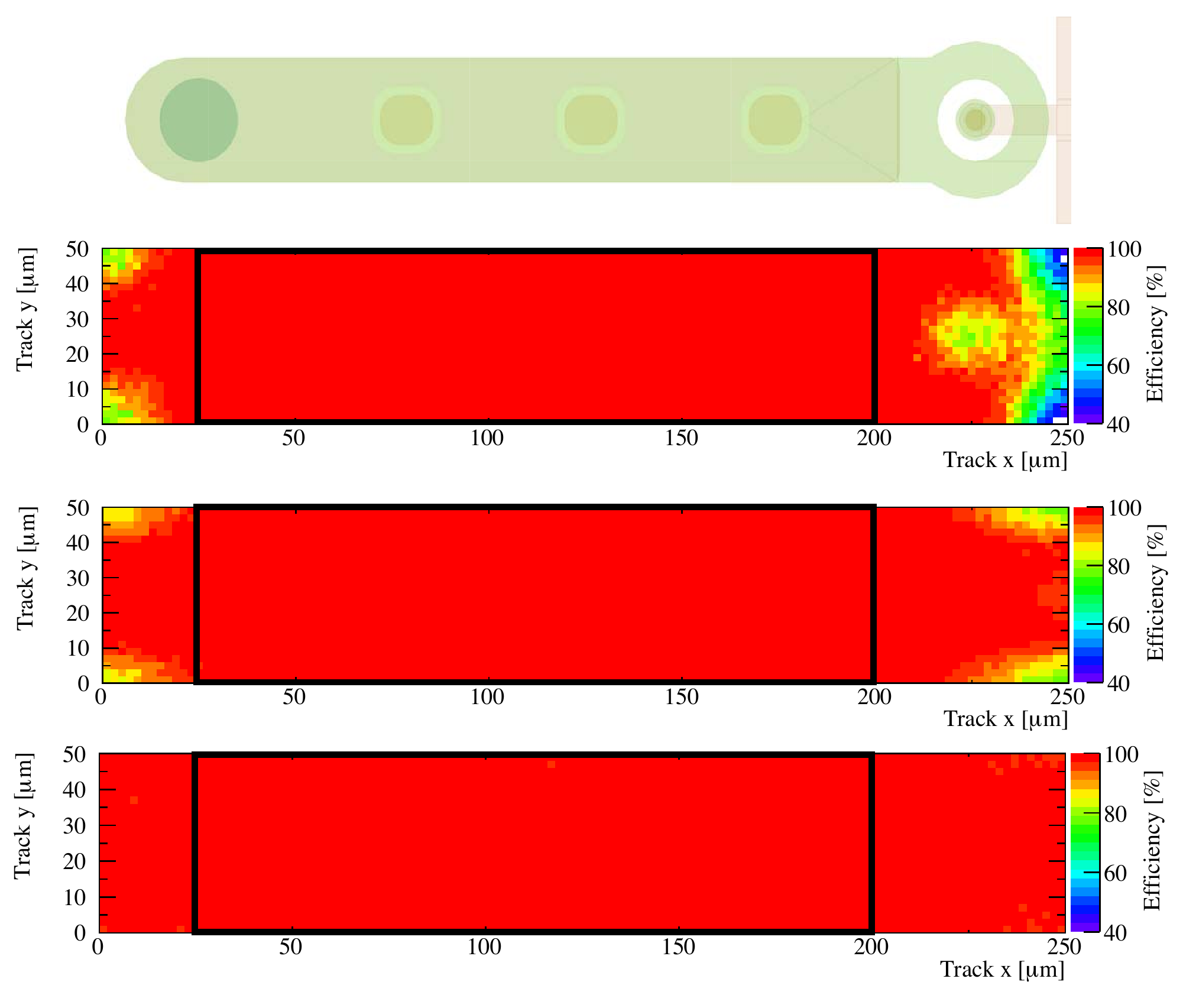}\label{fig:HLL-hieta-b}
}
\caption{Efficiencies of FE-I4 pixel modules with a 150 \mum{} thin sensor at different $\eta$. In \protect\subref{fig:HLL-hieta-a} the hit efficiency of the full pixel (filled dots) is compared to the inner pixel cell region (open dots) defined by the black rectangles in \protect\subref{fig:HLL-hieta-b}, where the hit efficiency over the single pixel surface is shown (from top to bottom) at 0$^{\circ}$ ($\eta$=0), 30$^{\circ}$ ($\eta\sim$0.55) and 45$^{\circ}$ ($\eta\sim$0.88) beam incident angle. }
\label{fig:HLL-hieta}
\end{figure*}

\begin{figure*}[tbp] 
\centering
\subfigure[]{
	\includegraphics[width=.47\textwidth]{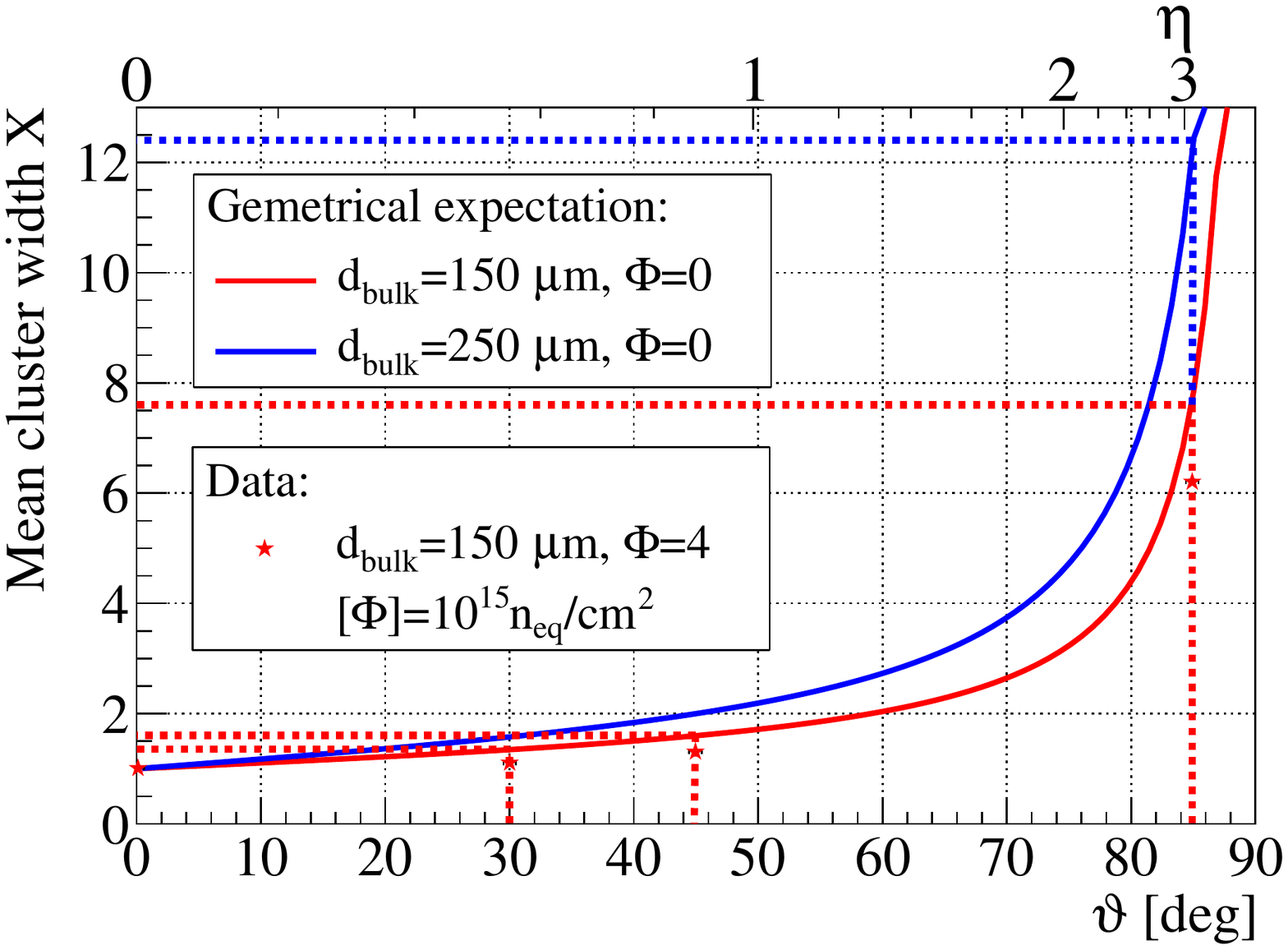}
	\label{fig:HLL-he-widthx_a}
}
\subfigure[]{
	\includegraphics[width=.48\textwidth]{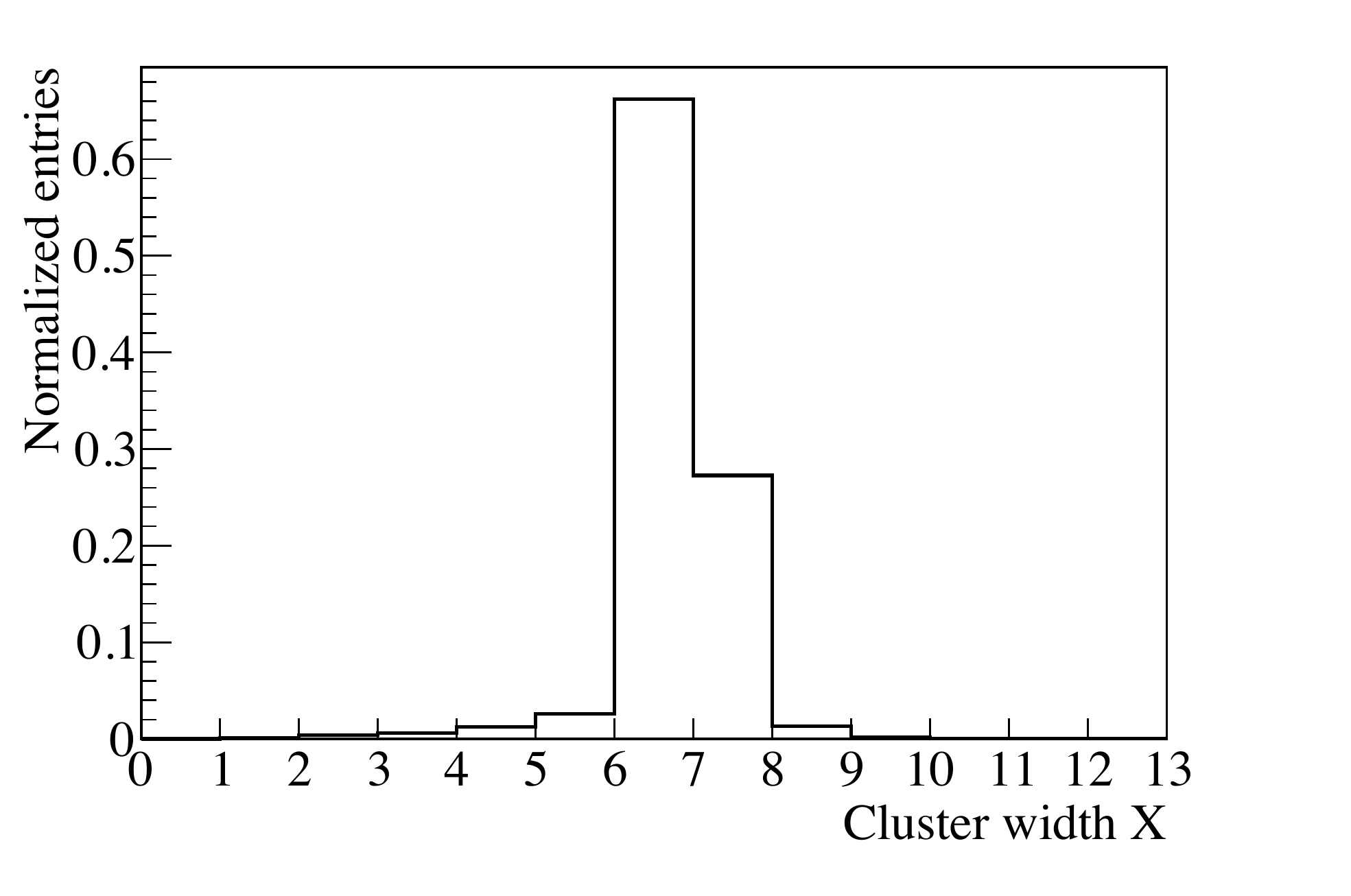}
	\label{fig:HLL-he-widthx_b}
	}
\caption{\protect\subref{fig:HLL-he-widthx_a} shows the mean cluster width along the tilted direction for different beam incident angles. For comparison the curves of the geometrical expectation for a not irradiated and fully depleted FE-I4 device with 150 \mum{} and 250 \mum{} thick sensors are also shown. The red stars denote beam test data from an FE-I4 150 \mum{} thin sensor after a fluence of 4$\times$10$^{15}$ \neqcm{} operated at $V_{\mathrm{bias}}=500$ V with a threshold of 1.6 ke. In (b) the cluster width distribution along the tilted direction for tracks at 85$^{\circ}$ incident angle with respect to normal incidence is shown.}
\label{fig:HLL-he-widthx}
\end{figure*}

\begin{figure*}[tbp] 
\centering
\includegraphics[width=1\textwidth]{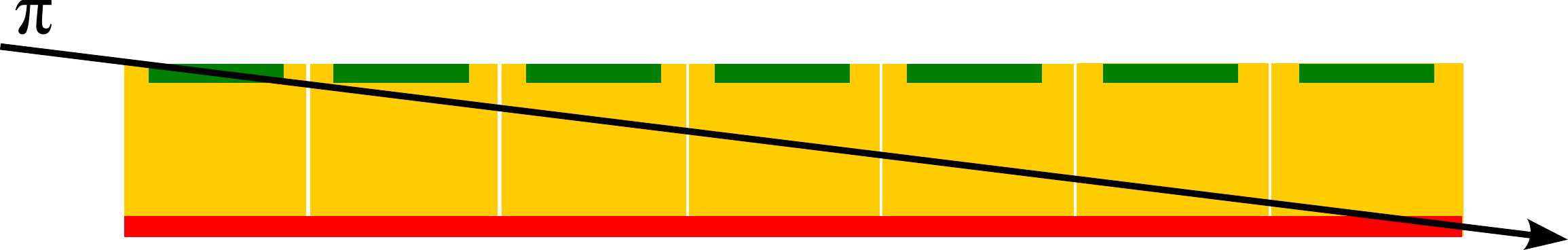}
\includegraphics[width=1\textwidth]{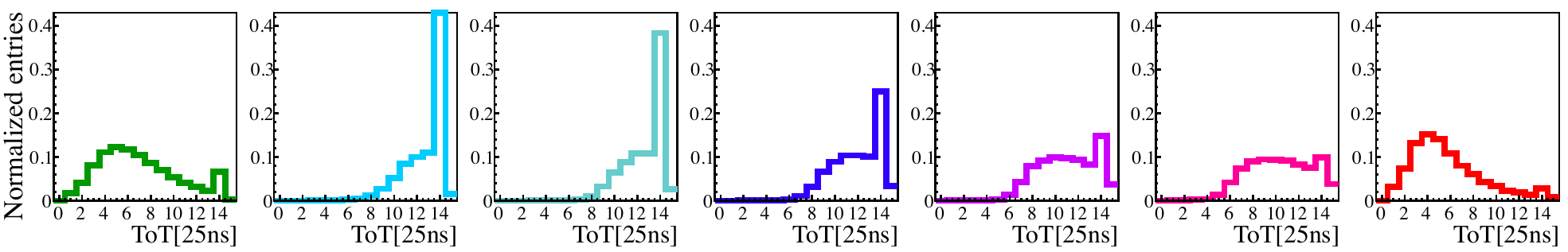}
\caption{The collected charge as a function of the Time over Threshold (ToT) for each cell of a 7 pixel long cluster. From left to right the particle is passing the module with increasing distance to the pixels on the module surface. In case of charge sharing along the short pixel side, the charge shown corresponds to the the sum of the two adjacent pixels. The 14th bin encodes the overflow charge above the calibration range.  Depending on the particle entrance point on the left most pixel, the two pixels at the edge of the cluster are only partially crossed resulting in a lower collected charge.}
\label{fig:HLL-he-tot}
\end{figure*}

\paragraph{Eta analysis.}
The behavior at different pseudo-rapidities has also been studied by tilting the detector along the axis perpendicular to the long pixel side up to 85$^{\circ}$ ($\eta\sim$3.1). For these studies the detector has been operated at $V_{\mathrm{bias}}=500$ V and with a threshold of 1.6 ke. Results are reported in figure~\ref{fig:HLL-hieta}, where the hit efficiency over the full pixel cell is compared to the one of the central part excluding the bias structures and the four pixel corners. The efficiency increases with the track incident angle up to a homogenous hit efficiency over the full pixel cell of 99.5\% at 45$^{\circ}$. A characterization of the pixel modules at 85$^{\circ}$ track incidence ($\eta\sim$3.1) has been performed to understand their behavior at the edge of the innermost layer of the planned HL-LHC detector. In this configuration a particle is expected to cross almost 1.7 mm inside a 150 \mum{} thick fully depleted bulk region. For the FE-I4 pixel shape this leads to a mean cluster width in the tilted direction of 7.4 pixels. Compared to the thickness of the present ATLAS pixel sensors of 250 \mum{}, for which in the same conditions a mean cluster width of 12.4 cells is expected, the thinner bulk has the advantage of reducing the occupancy at high-eta as demonstrated in figure~\ref{fig:HLL-he-widthx_a}. After irradiation the observed cluster width distribution along the tilted direction in figure~\ref{fig:HLL-he-widthx_b} shows a mean of 6.2 pixels and the corresponding hit efficiency is compatible with 100\% as expected from the hit efficiency definition. The discrepancy of the cluster size with respect to the pure geometrical expectation for a not irradiated module in figure~\ref{fig:HLL-he-widthx_a} is mainly due the not fully depleted bulk at 500 V and the effect of the threshold on partially crossed pixels at the edge of the long clusters. The effect of charge trapping as a function of the collecting distance is investigated by looking at the deposited charge inside the various pixels of the cluster along the tilted coordinate. Comparing the collected charge in the central pixels of the cluster in figure~\ref{fig:HLL-he-tot}, where particles cross the same amount of silicon, a decrease of the signal is observed as particles cross the depleted bulk farther from the implants. 

\subsection{Thin active edge pixels produced at VTT}\label{sec:active}
\begin{figure*}[tbp] 
\centering
\subfigure[]{
	\includegraphics[width=.48\textwidth]{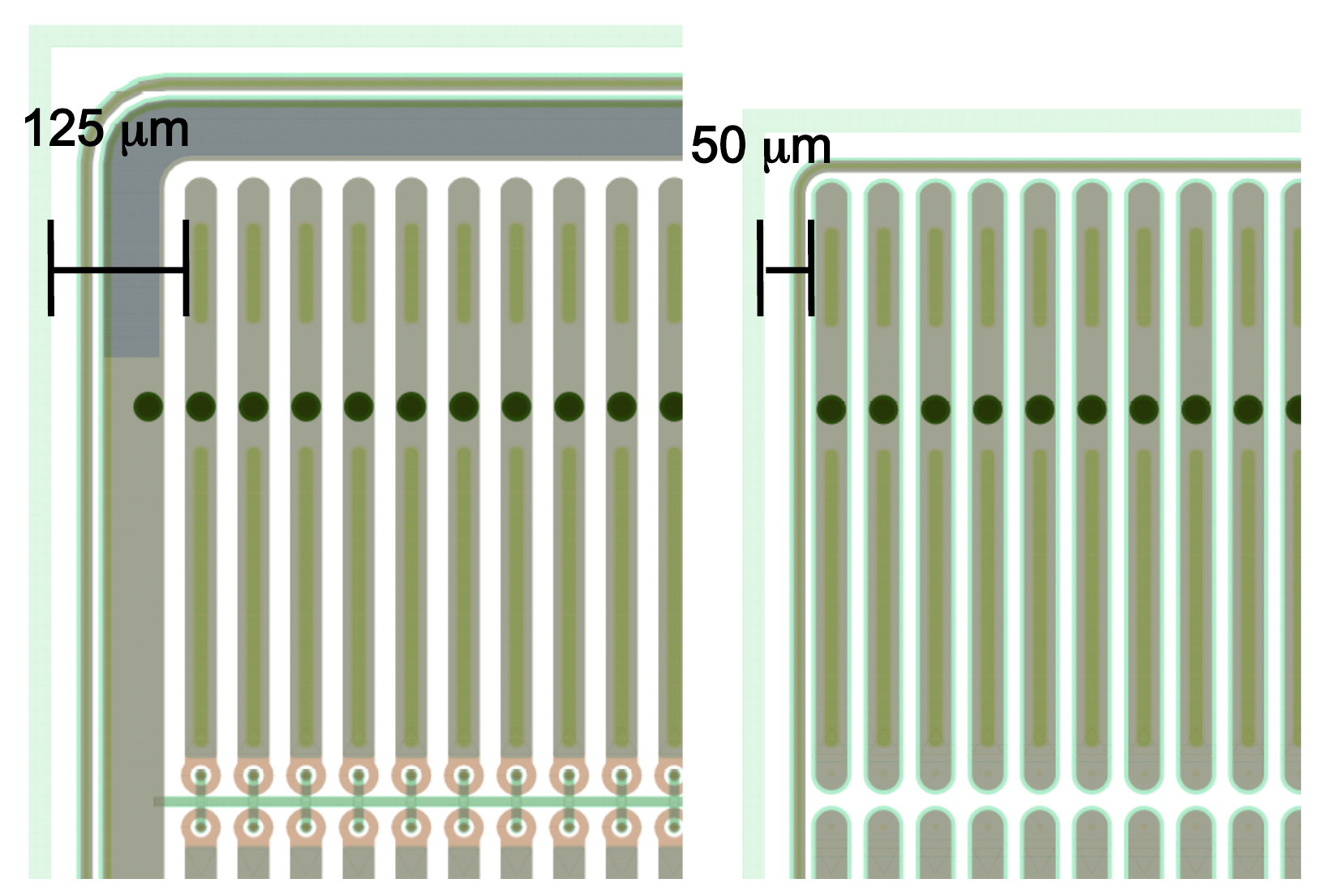}
	\label{fig:VTT-concept_edges}
}
\subfigure[]{
	\includegraphics[width=.475\textwidth]{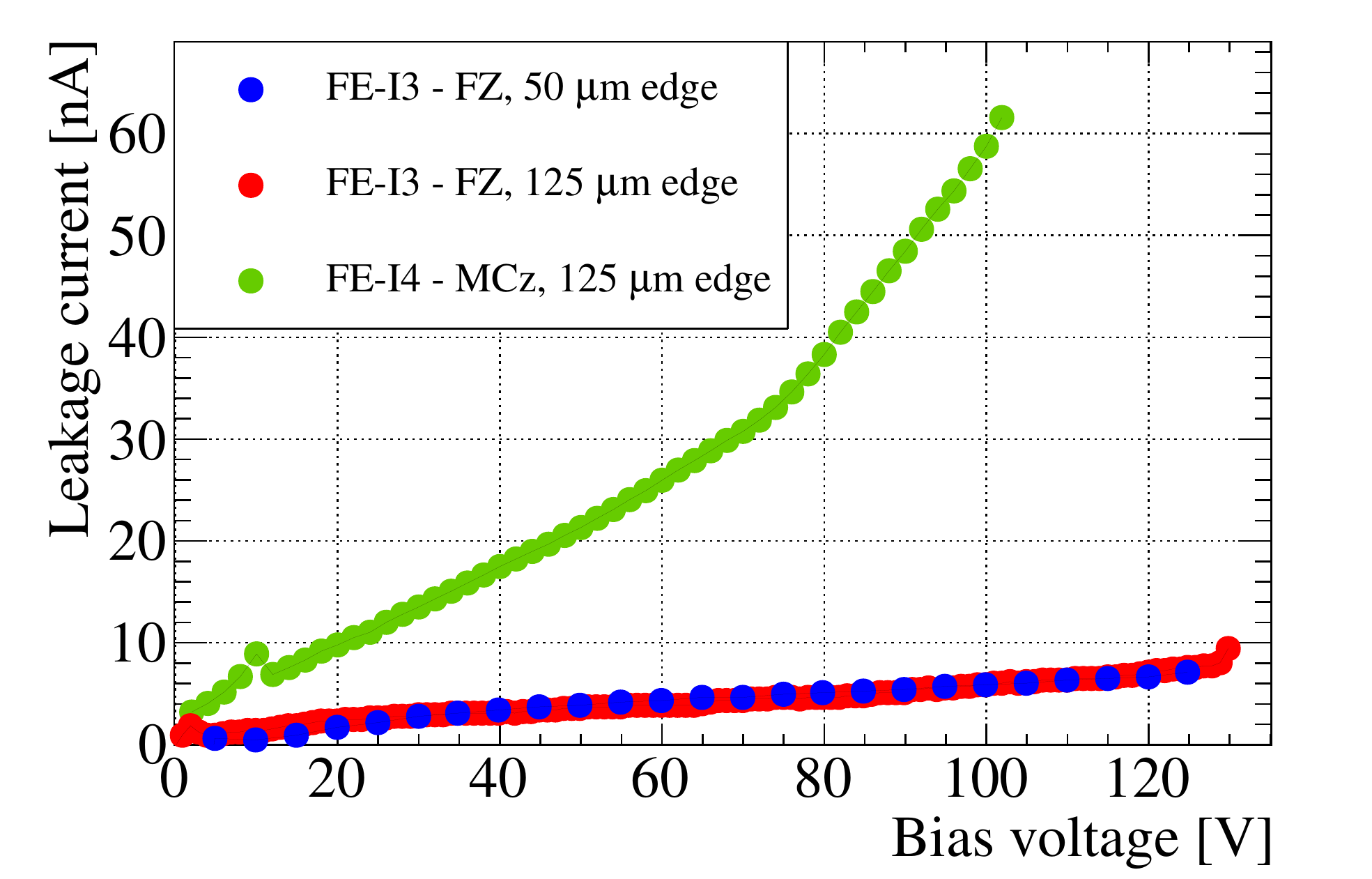}
	\label{fig:VTT-concept_iv}
}
\caption{\protect\subref{fig:VTT-concept_edges} Top view of the two different edge designs of the VTT sensors. The edge design on the left features a bias ring with a floating guard ring, and a distance from the last pixel implant to the sensor border of 125 \mum{}. A more aggressive edge design is sketched on the right. It uses only a floating guard ring and has a reduced distance to the active edge of 50 \mum{}. \protect\subref{fig:VTT-concept_iv} IV curves of the tested 100 \mum{} thin VTT sensors with different edge structures, total area and bulk silicon.}
\label{fig:VTT-concept}
\end{figure*}

N-in-p planar pixel sensors with active edges have been produced at VTT Finland, on p-type FZ 
silicon with an initial resistivity of 10 k$\Omega\,$cm, and on Magnetic Czochralski (MCz) silicon with orientation $\langle$100$\rangle$ and initial resistivity of 2 k$\Omega\,$cm. The sensors are thinned to 100 \mum{} and 200 \mum{}. The fabrication requires a support wafer that allows for etching trenches at the sensor borders using Deep Reactive Ion Etching (DRIE). The boron implant present on the backside of the p-type sensors is then extended to the sides with a four-quadrant ion implantation~\cite{fqii1,fqii2}. Afterwards, the handle wafer is removed and the sensors are interconnected with solder bump bonding to either FE-I3~\cite{fei3} or FE-I4 ATLAS chips. Homogeneous p-spray has been used for the inter-pixel isolation. Two different slim edge designs have been implemented characterized by a distance between the last pixel implant and the sensor border of 50 \mum{} with just one floating guard ring, or a 125 \mum{} distance when employing also a bias ring structure. The IV curves before irradiation are shown in figure~\ref{fig:VTT-concept}: breakdown voltages are between 100 V and 130 V and the full depletion voltage is around 15 V as expected from the high bulk resistivity. These modules have been irradiated at KIT up to a fluence of 5$\times$10$^{15}$ \neqcm{} and results obtained after irradiation are discussed in section~\ref{sec:CC}.

\begin{figure*}[tbp] 
\centering
\subfigure[]{
	\includegraphics[width=.47\textwidth]{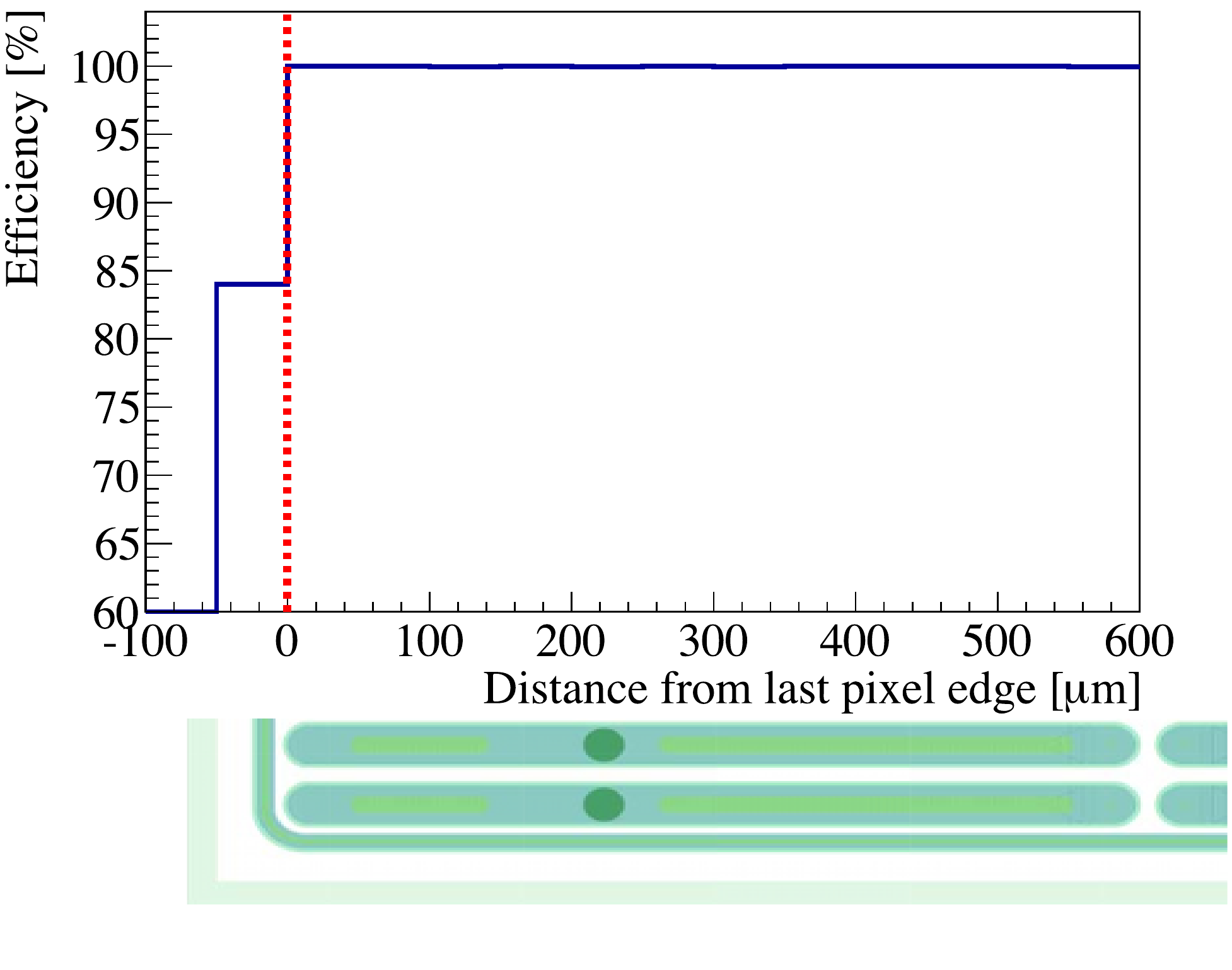}
	\label{fig:VTT-edge-50-a}
}
\subfigure[]{
	\includegraphics[width=.429\textwidth]{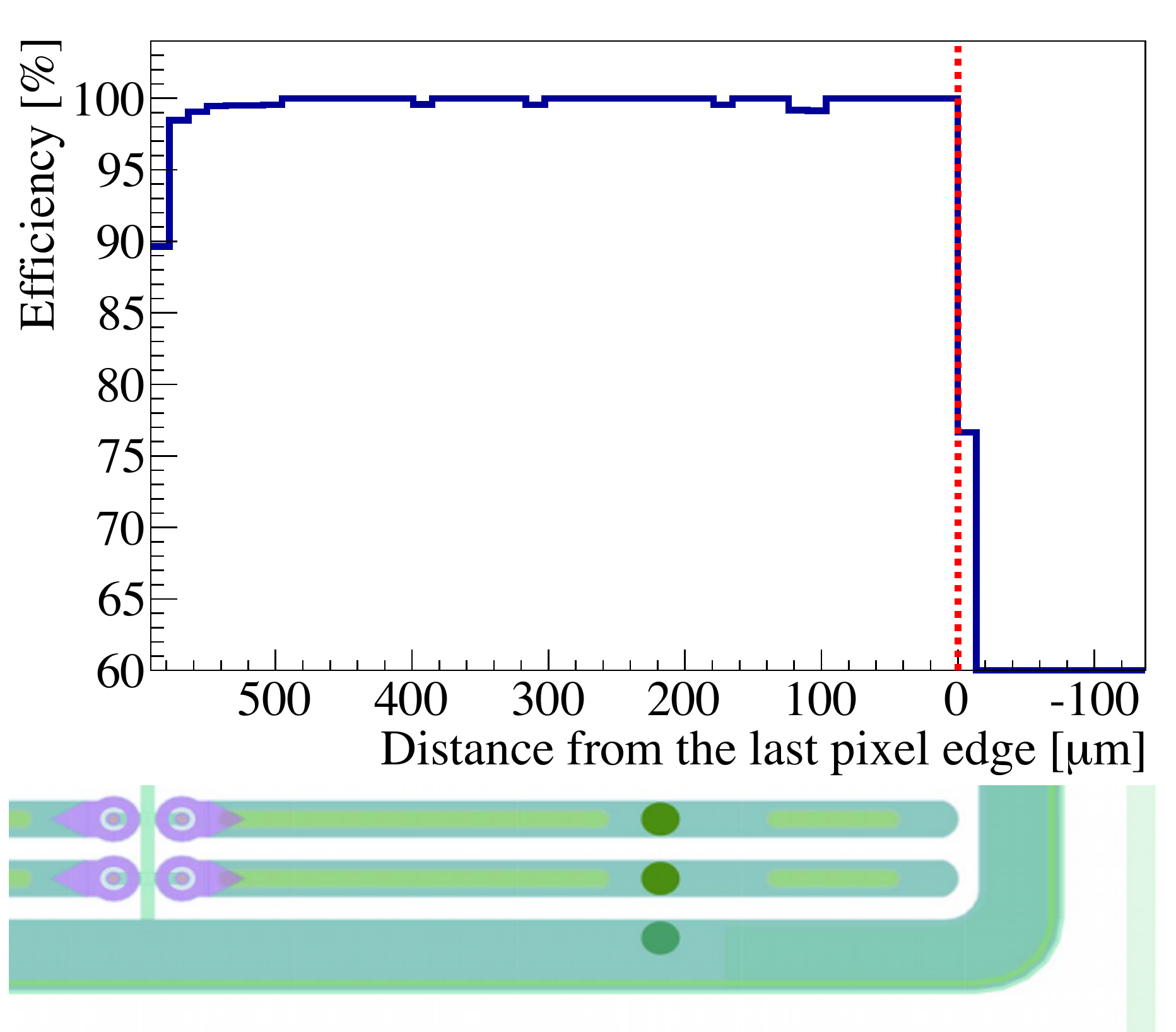}
	\label{fig:VTT-edge-50-b}
}
\caption{Efficiency of the last pixel column up to the active edge of FE-I3 100 \mum{} thin sensors. The red line denotes the end of the pixel implant: positive values of the X axis are relative to the last pixel implant while negative values indicate the border area outside the last pixel implant. \protect\subref{fig:VTT-edge-50-a} Left column of the 50 \mum{} edge design without bias ring. \protect\subref{fig:VTT-edge-50-b} Right column of the 125 \mum{} edge design with bias ring.}
\label{fig:VTT-edge-50}
\end{figure*}

\paragraph{Edge analysis.} Measurements of the hit efficiency over the sensor edge before irradiation have been performed at the CERN SpS facility with a 120 GeV beam of perpendicular incident pions. The experimental setup described in section~\ref{sec:thin} has been optimized to maximize the statistic for one border column for each of the two different slim edge designs.
The global efficiency for the inner pixels is more than 99.8\% for both modules. In figure~\ref{fig:VTT-edge-50} the hit efficiency is shown as a function of the distance from the last pixel column for the two edge structures of 100 \mum{} thin FZ sensors. For the 50 \mum{} edge design a particle traversing this border area can be detected on the last pixel implant with a measured hit efficiency of 84$^{+9}_{-14}$\%. In contrast, for the 125 \mum{} edge design the charge created beyond the last pixel implant is partially or totally collected by the bias ring structure. Therefore a lower hit efficiency of 77$\pm$1\% is observed only in the 15 \mum{} between the last pixel implant and the bias ring.

\subsection{Charge collection studies after irradiation}\label{sec:CC}
\begin{figure*}[tbp] 
\centering
\subfigure[]{ \includegraphics[width=.45\textwidth]{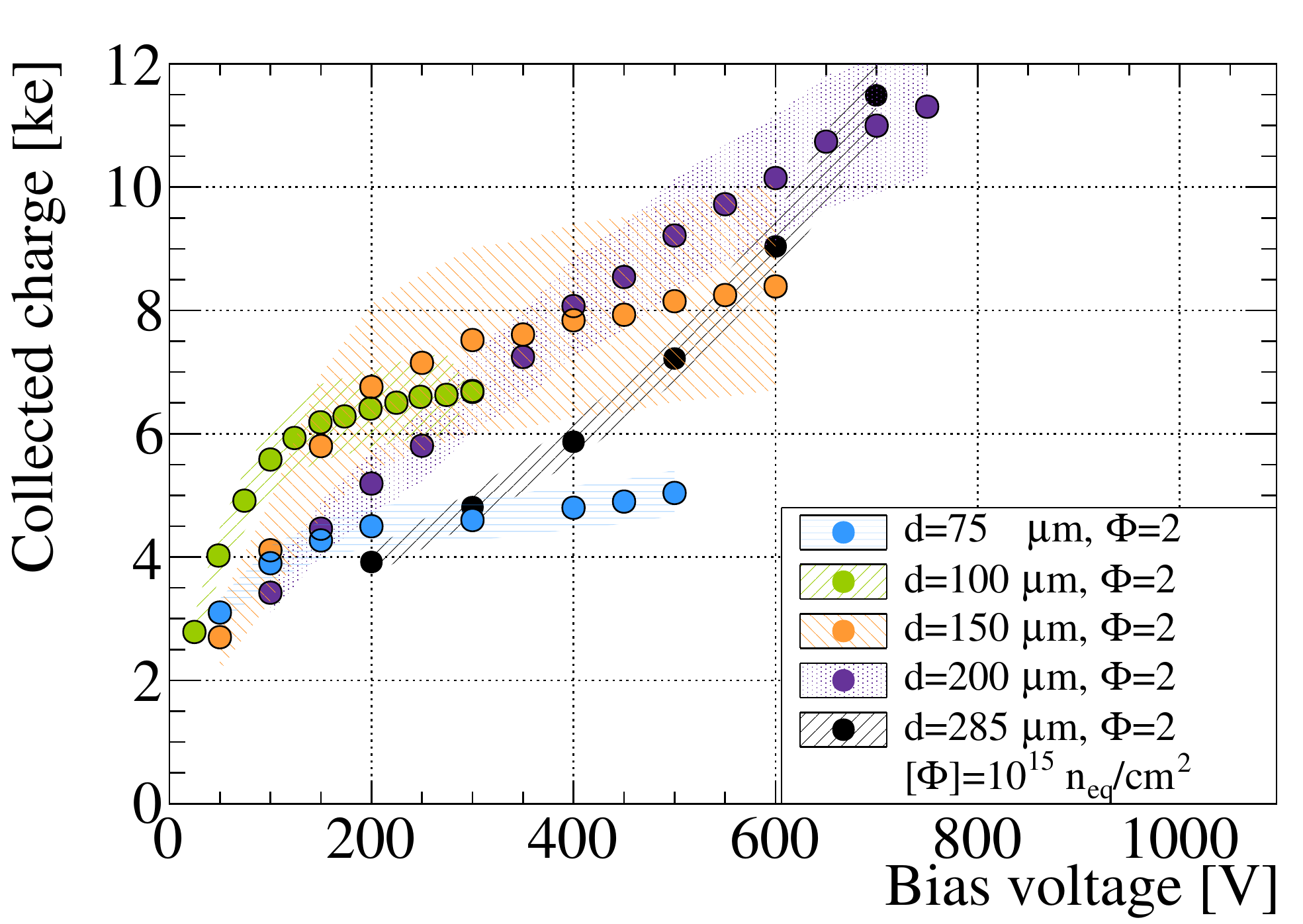} \label{fig:thick-comp-2e15} }
\subfigure[]{ \includegraphics[width=.45\textwidth]{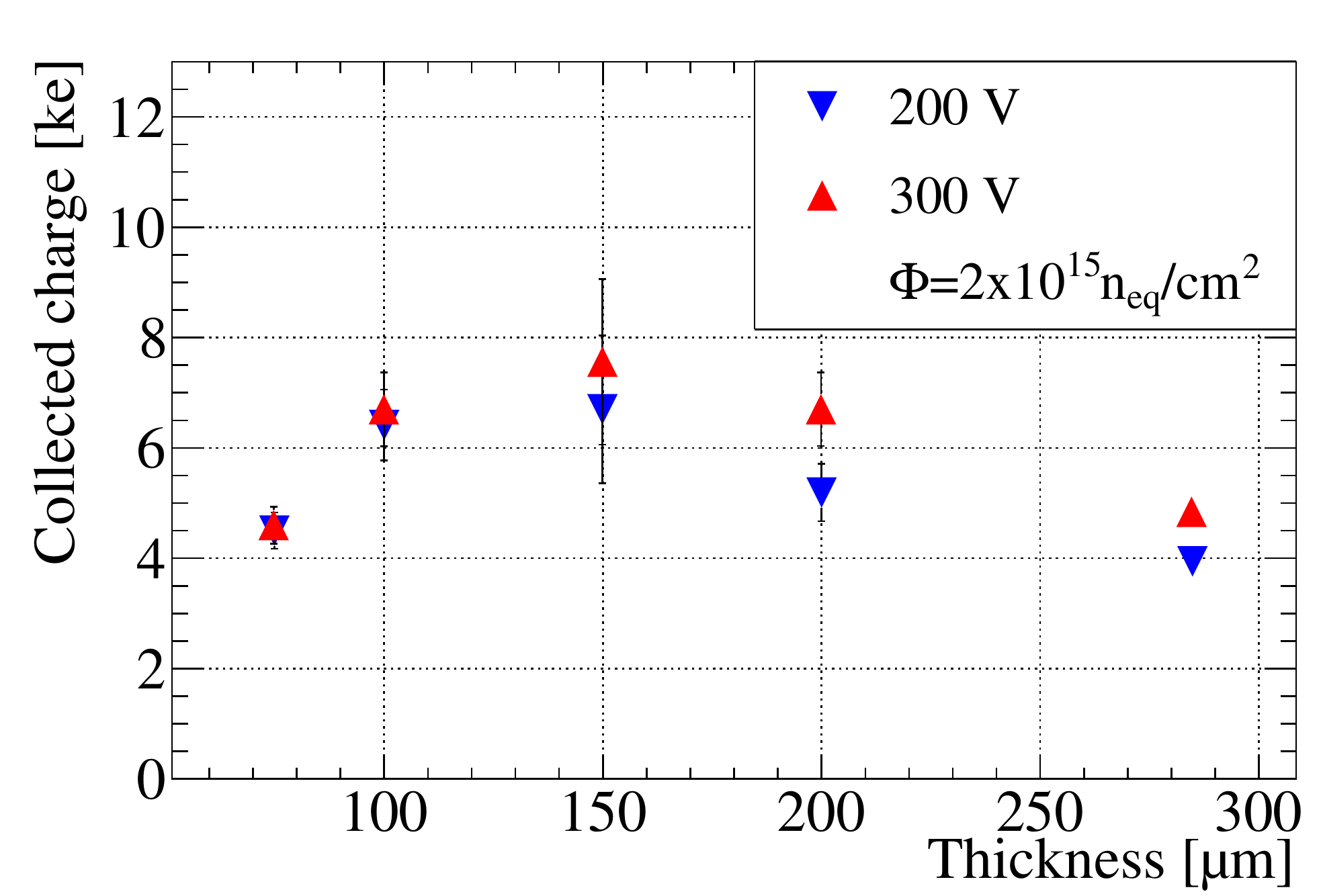} \label{fig:thick-comp-2e15-MPV} }
\subfigure[]{ \includegraphics[width=.45\textwidth]{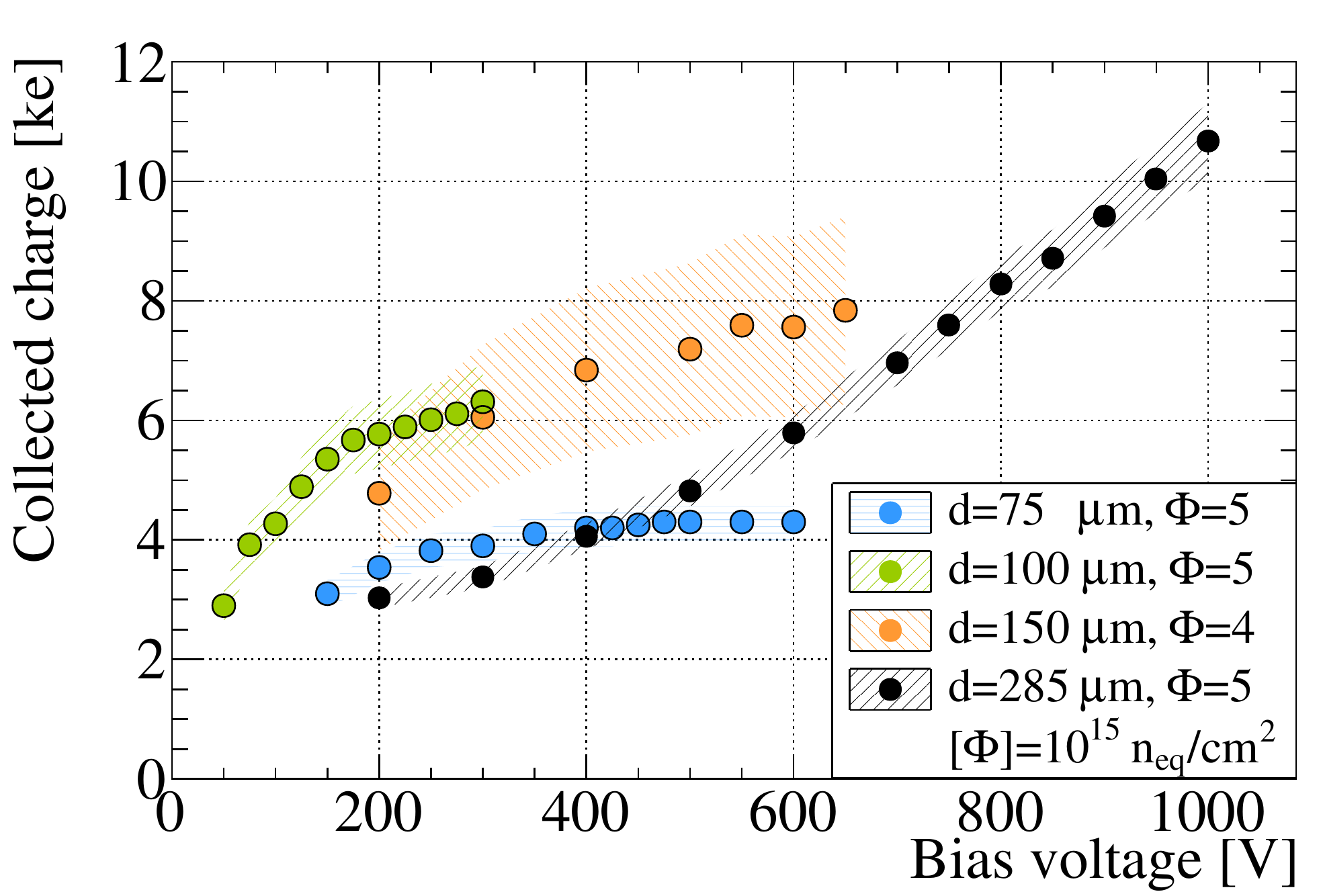} \label{fig:thick-comp-5e15} }
\subfigure[]{ \includegraphics[width=.45\textwidth]{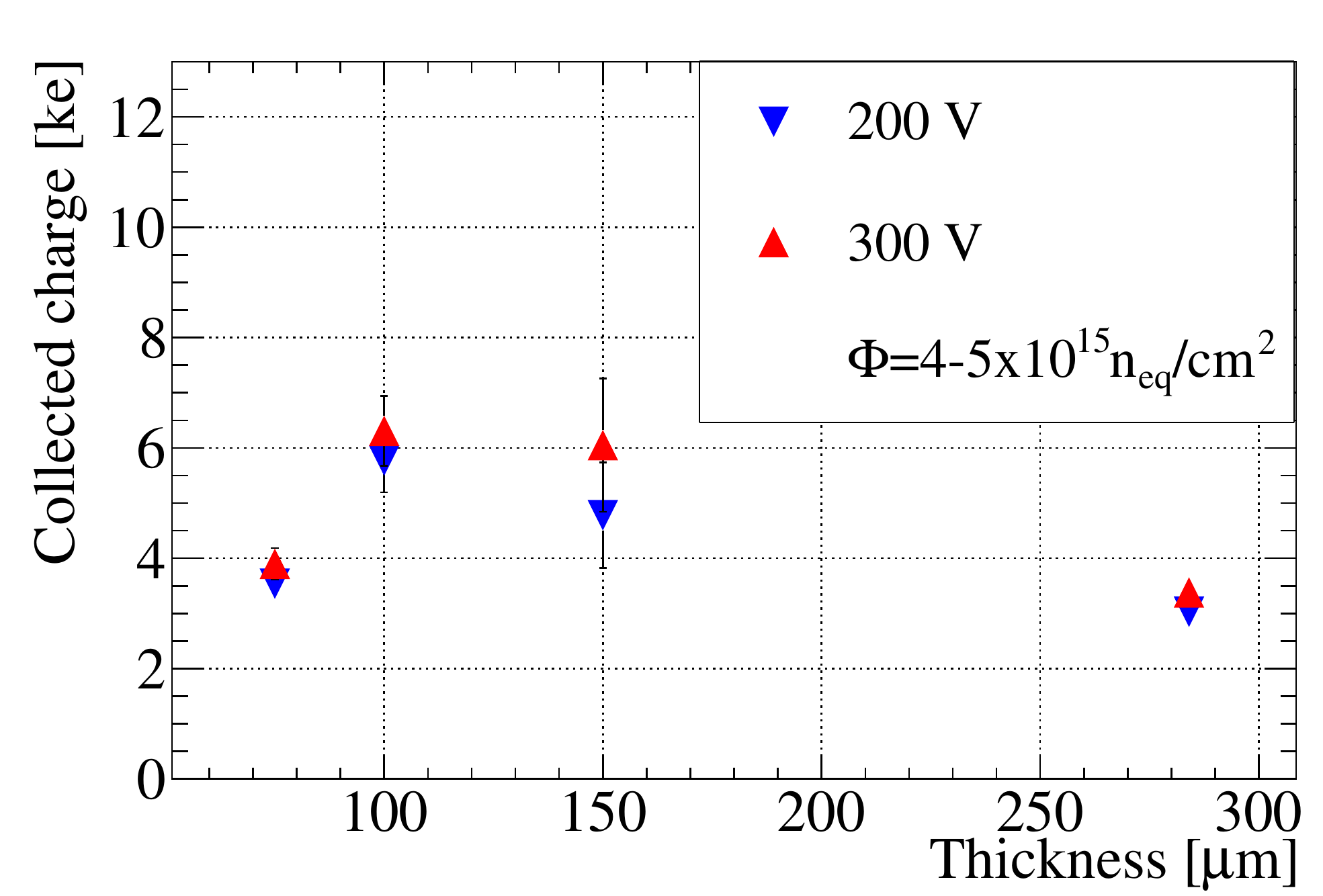} \label{fig:thick-comp-5e15-MPV} }
\subfigure[]{ \includegraphics[width=.45\textwidth]{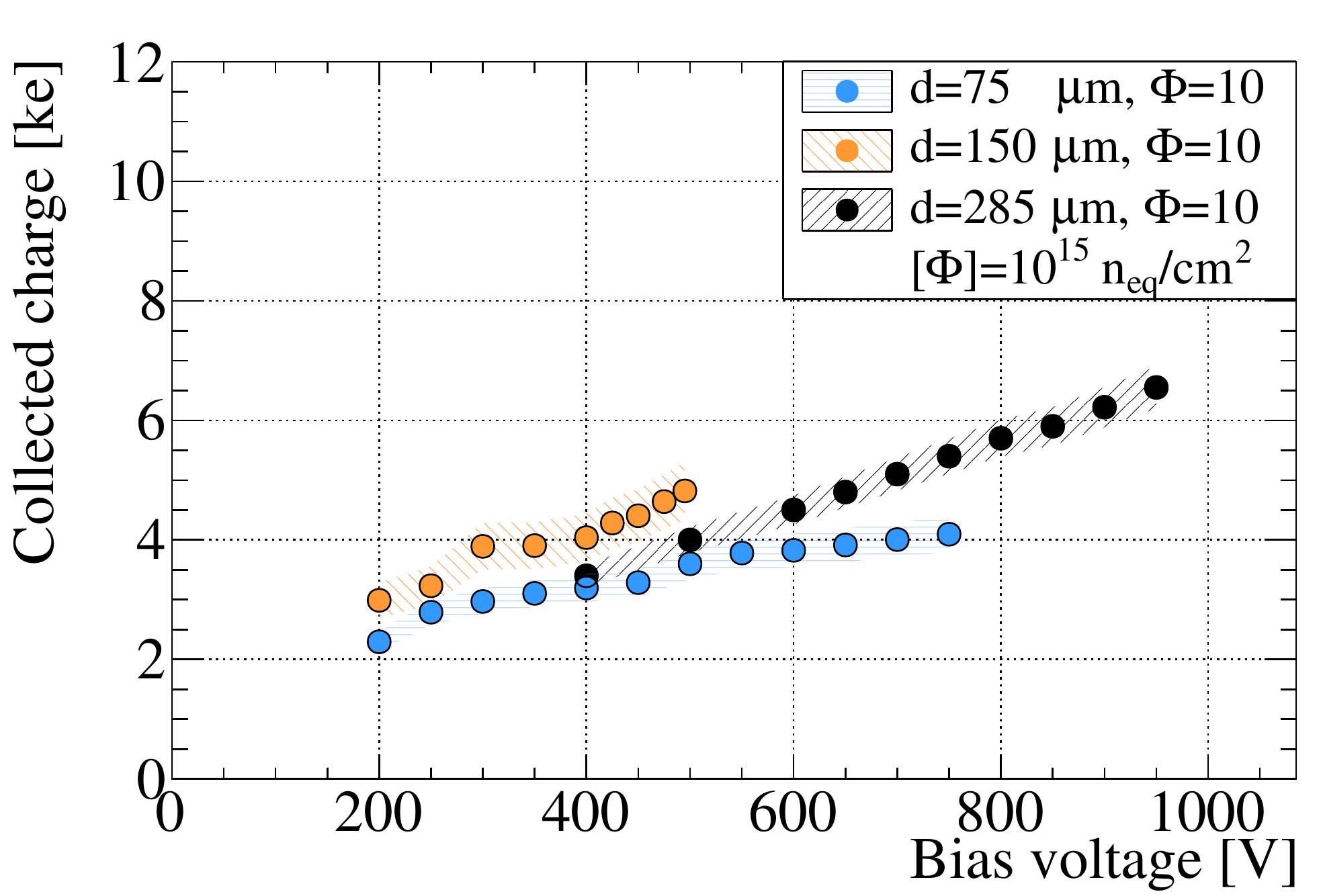} \label{fig:thick-comp-1e16} }
\caption[]{Overview of the charge collection for irradiated pixel modules of different thicknesses. The results after a fluence of 2$\times$10$^{15}$ \neqcm{}, 4-5$\times$10$^{15}$ \neqcm{} and 10$^{16}$ \neqcm{} are shown in \protect\subref{fig:thick-comp-2e15}, \protect\subref{fig:thick-comp-5e15}, \protect\subref{fig:thick-comp-1e16}, respectively. The collected charges at 200 V and 300 V for the different sensor thicknesses compared in \protect\subref{fig:thick-comp-2e15} and \protect\subref{fig:thick-comp-5e15} are highlighted in \protect\subref{fig:thick-comp-2e15-MPV} and \protect\subref{fig:thick-comp-5e15-MPV}, respectively.
}
\label{fig:thick-comp}
\end{figure*}

Studies of the charge collection for the presented structures have been performed in the laboratory using electrons from a \Sr{} beta source triggered with an external scintillator.
To be consistent with beam test measurements the full setup is kept at a constant temperature in a climate chamber with an environmental temperature of -50\degC{} (corresponding to about -40\degC{} on the chip~\cite{weigell}). For the read-out the USBPix system is used. Before irradiation, the sensors show the expected collected charge after the full depletion of about 7 ke and 12 ke for thicknesses of 100 \mum{} and 150 \mum{}, respectively~\cite{slid-tsv}. In figure~\ref{fig:thick-comp} results after irradiation of the 100 \mum{} thin MCz from VTT and the 150 \mum{} thin sensors from the MPP/HLL production are compared to measurements of 75 \mum{}, 200 \mum{} and 285 \mum{} thick devices obtained in previous studies~\cite{n-in-p,weigell,slid-tsv}. Whenever possible, the systematic uncertainty on the ToT to charge calibration has been reduced from 20\% to 10\% using gamma radiation from \Am{} and \Cd{} sources as reference. After irradiation, thin sensors up to 150 \mum{} approach the full depletion with moderate voltages between 200 and 300 V. In this voltage range 100 \mum{} and 150 \mum{} thin sensors appear to be the best compromise between active volume and trapping effects resulting in the highest collected charge after a fluence up to 4-5$\times$10$^{15}$ \neqcm{} (see figure~\ref{fig:thick-comp-2e15-MPV} and \ref{fig:thick-comp-5e15-MPV}). At the highest measured fluence of 10$^{16}$ \neqcm{}{} trapping is the dominating effect and the collected charge of the 150 \mum{} thin sensors becomes similar to the one of the 75 \mum{} and 285 \mum{} thick sensors up to 500 V.

\section{Conclusions}\label{sec:conclusions}
The characterization of n-in-p planar pixel devices employing 150 \mum{} thin pixel sensor and 100 \mum{} thin active edge sensors has been presented. At moderate voltages (between 200 and 300 V) up to a fluence of 4-5$\times$10$^{15}$ \neqcm{} their charge collection after irradiation is superior to those measured for sensors of other thicknesses ranging from 75 \mum{} to 285 \mum{}. After a fluence of 10$^{16}$ \neqcm{}, 75 \mum{} 150 \mum{} and 285 \mum{} thick sensors show similar collected charge up to a bias voltage of 500 V.
The hit efficiency of the n-in-p 150 \mum{} thin sensors has been measured for tracks with perpendicular incidence up to a fluence of 4$\times$10$^{15}$ \neqcm{}, where a hit efficiency of 97.7\% is obtained at the bias voltage of 690 V. The inefficiency in the punch through and the bias rail structures is reduced for inclined tracks up to a homogeneous efficiency distribution over the full pixel cell of 99.5\% with a track incident angle of 45$^{\circ}$. The good performance of active edge sensors with both 125 \mum{} slim edge and 50 \mum{} active edge produced at VTT has been demonstrated showing a hit efficiency even in the 50 \mum{} outside the last pixel implant if no bias structure is present.

\acknowledgments
This work has been partially performed in the framework of the CERN RD50 Collaboration. The authors thank A.~Dierlamm (KIT), S.~Seidel (NMU), V.~Cindro, and I.~Mandic (Jo\v{z}ef-Stefan-Institut) for the sensor irradiations. Part of the irradiation were supported by the Initiative and Networking Fund of the Helmholtz Association, contract HA-101 ("Physics at the Terascale"). Another part of the irradiations and the beam test measurements leading to these results has received funding from the European Commission under the FP7 Research Infrastructures project AIDA, grant agreement no. 262025. Beam test measurements were conducted within the PPS beam test group.

\end{document}